# Reflections on the Usefulness and Limitations of Tools for Life-Centred Design

Martin Tomitsch[a,b], Katharina Clasen[b], Estela Duhart[b], Damien Lutz[b]

[a]Transdisciplinary School, University of Technology Sydney, Australia

[b]Life-centered Design Collective

*Corresponding author e-mail: martin.tomitsch@uts.edu.au

*All authors contributed equally to the study, the first author led the paper writing process, the other authors are listed in alphabetical order.*

**Abstract**: Life-centred design decenters humans and considers all life and the far-reaching impacts of design decisions. However, little is known about the application of life-centred design tools in practice and their usefulness and limitations for considering more-than-human perspectives. To address this gap, we carried out a series of workshops, reporting on findings from a first-person study involving one design academic and three design practitioners. Using a popular flat-pack chair as a case study, we generatively identified and applied four tools: systems maps, actant maps, product lifecycle maps and behavioural impact canvas. We found that the tools provided a structured approach for practising systems thinking, identifying human and non-human actors, understanding their interconnectedness, and surfacing gaps in the team's knowledge. Based on the findings, the paper proposes a process for implementing life-centred design tools in design projects.

## 1. Introduction

Organisations embrace human-centred design for its ability to deliver things that people want (Borthwick et al., 2022). But what people want is not always good for the planet and has ripple effects on communities and ecosystems (Monteiro, 2019). Human-centred design emerged from the deep desire to make technology work better for people (A. Cooper, 2004; Giacomin, 2015), but in many instances, it has evolved into an instrument for accelerating business outcomes and growth (Sheppard et al., 2018). Driven by the neoliberal economic system, organisations have employed human-centred design, working with agencies and building in-house capabilities to outperform competitors through creating human-centric products, services and systems.

Centring innovation on the needs and desires of users and consumers risks omitting environmental and ethical concerns. To meet sustainability targets, organisations instead rely on compensation strategies and post-processing activities (Demirel & Duffy, 2013). Even Apple's first carbon-neutral smartwatch released in September 2023 offsets some of the carbon emissions through environmental initiatives like restoring forests. The literature defines human-centred design in ways that consider inclusive, diverse and sustainable perspectives (ISO, 2019; Krippendorff, 2004). However, how the approach is adopted in practice, often is limited and narrow (A. Cooper, 2004; Lutz, 2022; Monteiro, 2019).

While sharing many fundamental principles with human-centred design, the notion of life-centred design emphasises a shift towards considering all living beings and the far-reaching impacts of design decisions (Borthwick et al., 2022; Forlano, 2016; Lutz, 2022). Designing through this lens eliminates the need for offsetting carbon emissions associated with production, use and disposal. Instead, life-centred tools provide a mechanism to uncover these impacts during the design phase and to identify alternative approaches that create a systemic balance.

This paper focuses specifically on the usefulness and limitations of design tools that support a more systemic and sustainable approach with the aim of creating a balanced economy in which all living beings, communities and ecosystems can thrive. Based on a review of design tools, we generatively arrived at four tools that we applied in our study: systems maps, actant maps, product lifecycle maps and behavioural impact canvas. To provide an initial investigation, we used the tools to systematically identify the environmental and social impacts of IKEA's bestselling Poäng chair and generate alternative solutions.

In the following sections, we briefly outline the evolution of human-centred design approaches and the emergence of life-centred design. We then describe our methodology, which includes the process of using the tools and analysing the data collected through our first-person study. This is followed by a description of the tools, an analysis of the workshops and a discussion of our findings.

The paper makes two specific contributions. Firstly, it presents a documented account of applying life-centred design tools for the purpose of reviewing an existing product. Secondly, we present a process for getting started with life-centred design in design education and practice and discuss how the tools support achieving the life-centred objectives.

## 2. Background

For over a century, organisations have harnessed the skills of designers as a source of innovation and to achieve a competitive advantage (Owen, 1990). Human-centred design emerged in response to the pace of technological development. Instead of starting with a technology looking for a solution, human-centred design begins with identifying the needs of users or consumers (A. Cooper, 2004). This includes a focus on stakeholders, context of use and creative processes (Maguire, 2001). To implement this approach, designers apply tools and collect data for answering questions that “span the spectrum from the physical nature of people’s interaction with the product, system and service to the metaphysical” (Giacomin, 2015).

In recent years, human-centred design has been criticised for its anthropocentrism, which prioritises the needs, desires and wellbeing of a select group of human actors above others that are marginalised, non-human and do not have a voice in the design process (Costanza-Chock, 2020; Norman, 2023). The approach has also received criticism for its focus on shareholder primacy, using human-centred design as a way to monetise consumers, and contributing to the depletion of resources and rate of growth in consumer products with short lifespans (T. Cooper, 2017; Foth et al., 2021; Monteiro, 2019). These limitations of human-centred design have given rise to alternative approaches, labelled as life-centred, planet-centred, environment-centred, country-centred, environmentally sustainable, value-sensitive, more-than-human or regenerative design. What all these approaches have in common is a call for consciously considering living systems when making design decisions.

In this paper, we use the term “life-centred” as it encapsulates the idea of decentering humans and considering the needs of all, human and non-human (Lutz, 2022). Life-centred design tools facilitate this approach by expanding the human-centric focus and bringing systemic and far-reaching perspectives into the design process. Borthwick et al. (2022) drew on established principles from the field of environmental policy to guide the design of interactive products. Karpenja (2023) applied a life-centred design lens to the domain of product packaging, outlining how the field has evolved over time from a product-centred to a human-centred approach and is under increasing pressure to shift to a fully circular and waste-free process. El-Rashid et al. (2021) implemented a life-centred design approach to tackle the problem of space debris. Their process involved scenarios, roadmapping and backcasting as tools and their analysis found that life-centred design when applied at a systems level can evolve with the wicked problem to provide new perspectives.

The study presented in this paper builds on this previous body of work, contributing insights into the use and usefulness of life-centred design tools for moderating the impact that products, services and systems have on the environment and ecosystems.

## 3. Methodology

We report on a first-person study to collect data about using life-centred tools and reflect on the process and outcomes through a first-person perspective (Chang, 2013; Lucero et al., 2019). As such, the four authors of this paper represent the participants in the study and the work reported draws on data collected through a series of collaborative workshops and reflections. One of the authors works as a design academic at a university of technology, providing the academic framing for the study. The other three authors are practising designers who also engage in writing about design and teaching design at tertiary institutions. The authors live and work across three continents (Australia, North America, Europe), allowing us to draw on different cultural norms and practices. The combination of academic and practice-based perspectives represents an important aspect of the study and sets the study apart from other studies discussed in the background section.

Data was collected through six workshops held across the period of three months followed by a process of reflection and synthesis (Figure 1). Workshops took place fortnightly and employed Zoom as an online video conferencing tool and Miro as a digital collaboration board. Each session was recorded and auto-transcribed for subsequent analysis. All four authors participated in each of the workshops, except for the final workshop, which only involved three participants. The fourth participant watched the recording of the workshop and subsequently carried out the tool exercise in his own time, adding notes to the digital board. Each of the four workshops was led by one of the authors in rotation, which involved preparing templates and facilitating the online sessions. After the final workshop, each participant recorded their personal reflections using digital sticky notes responding to three prompts: What did we learn about the subject? What did we learn about the tool? What did we learn about the process? This was followed by another online session in which we compared and discussed our collective reflections. The second and fourth authors independently analysed the reflection notes through affinity mapping to identify common themes. The lead author then drew on both affinity maps to structure the findings. The third author reviewed the transcripts to link our reflections to quotes from the workshop sessions.

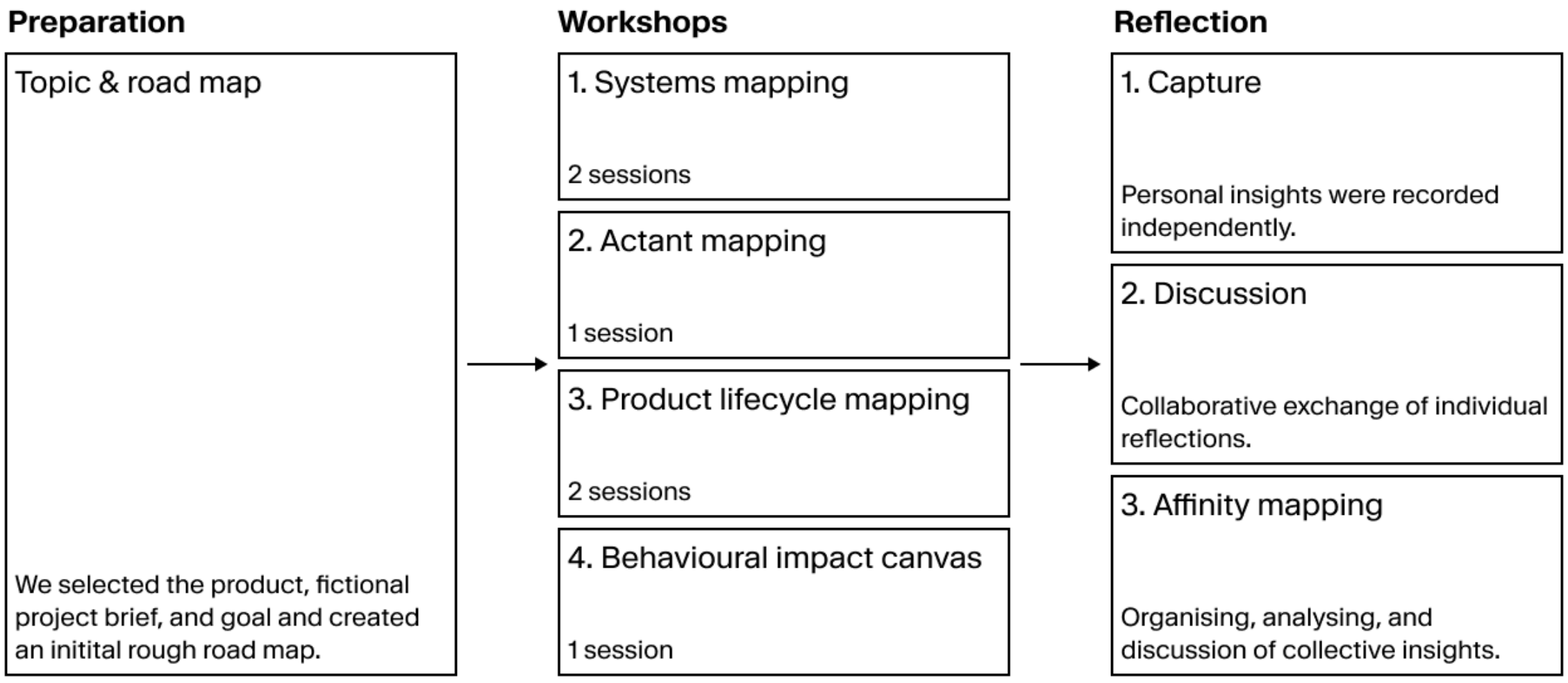

*Figure 1: The steps and activities carried out as part of the design process. Each workshop session lasted one hour and sessions were two weeks apart.*

As a first step, we brainstormed and recorded potential topics, including: a social media app, ChatGPT, electric vehicles, solar panels, a sofa/chair and virtual reality headsets. Following a voting process and a discussion about the pros and cons of each topic, we decided on the chair as a subject as it represents an everyday object that involves both physical as well as digital (i.e. online shopping) layers. To facilitate data collection, we narrowed our focus to a specific brand and model while creating a hypothetical yet realistic project brief. We selected IKEA's Poäng chair as it represents a top-selling product with a global footprint. IKEA engages in several environmental initiatives, such as sourcing wood from responsibly managed forests, extending the value of our study to companies that already pride themselves as leaders in sustainability.

As a second step, we decided on the tools to implement a life-centred design approach to redesigning the Poäng chair. As a source for this step, we drew on handbooks that include a focus on life-centred design (Lutz, 2022, 2023; Tomitsch, Borthwick, et al., 2021), previous studies outlined in the background section and the team's experience with using tools in their teaching. Instead of selecting all the tools at the start, we developed an initial rough roadmap for their sequencing. The initial roadmap (and tools) covered the phases of understanding/analysing the status quo (systems mapping, actant mapping), going deeper and focusing on the impacts (impact ripple canvas, sustainability strategy canvas) and focusing on a specific issue (non-human/non-user personas, journey mapping). Using this list, we started with systems mapping and subsequently chose each tool in a generative manner. The process of deciding on which tool to apply in each step drew on the authors' experience of using the tools in teaching, research and practice. The tools described in the next section are therefore only representative candidates of life-centred design tools chosen through the authors' perspective and lens.

## 4. The life-centred design tools

We decided to commence with systems mapping as a way to reveal some of the invisible components that could then be used as input for subsequent tools. Next, we developed an actant map to identify the human and non-human entities. This was followed by a product lifecycle map to visualise the natural and human resources used and impacted by the chair's lifecycle. As the fourth tool, we created a behavioural impact canvas to link human behaviour to the impacts uncovered by the previous tools and to brainstorm potential solutions. This section provides a detailed description of each tool.

### *4.1. Systems mapping*

Systems mapping is a method to support systems thinking in practice. Systems thinking was initially developed in the 1950s and applied to large-scale technological systems, such as aircraft, missile systems and nuclear power plants (Forrester, 1999). In the second half of the 20th century, the application of systems thinking was extended to social and organisational problems, such as organisational change, strategy development, and leadership (Meadows, 2008; Senge & Sterman, 1992). Systems maps are a tool to represent the components that make up a system. Directional arrows describe the effect that one component has on another.

This effect can be negative (a decrease) or positive (an increase). Systems maps also reveal feedback loops, which are either balancing or reinforcing. Mapping out the components and their relationships shows how events within a system or across multiple systems are connected and to assess the impact of interventions, for example, to increase the uptake of cycling (Macmillan et al., 2014) and restrict the spread of the COVID-19 virus (Sahin et al., 2020). As a life-centred tool, systems maps offer two qualities: First, they can reveal entities and systems beyond the obvious ones affected by an intervention. Second, they highlight the far-reaching impacts of an intervention (referred to as second and third order impacts). For our study, we employed an approach to systems mapping that starts with a mind map to identify potential components. As input, each participant gathered data about the Poäng chair from online sources and brought this data to the first workshop.

### *4.2. Actant mapping*

Actant mapping is a variation of stakeholder mapping (Sznel, 2020). Traditionally, stakeholder mapping is part of stakeholder analysis, a process that involves assessing people, groups or organisations that have an impact or can be impacted by a project or initiative. In design, stakeholder analysis helps understand needs, audience and level of influence. Actant mapping is based on the observation that this process in the past commonly failed to acknowledge non-humans as stakeholders (Sznel, 2020). Instead of stakeholders, it focuses on actants, drawing on Latour's definition that extends to non-human, non-individual entities (Latour, 1996) and the Actor-Network Theory, a sociological and philosophical framework that states that both society and knowledge are constructed by the relationships between both human and non-human "actors" – people, concepts, objects (Latour, 2005). For Latour, all actors have equal agency and can influence a situation but do not have meaning inherently; meaning comes from their relationships in the network (Latour, 2005). It is with this perspective that the actant mapping tool helps shift the perception of non-humans as resources towards actors with equal agency in the design process. As a life-centred design tool, actant maps encourage a more balanced view of the entities involved in or affected by a design intervention.

### *4.3. Product lifecycle mapping*

The product lifecycle map is a lifecycle analysis tool for mapping the natural and human resources used and impacted by a product's lifecycle (Lutz, 2023). It provides a systematic approach for investigating the complete lifecycle (from raw materials to final disposal) of products (Williams, 2009). The product lifecycle map implements this approach through a dedicated section for capturing what happens to the product and its materials, from sourcing materials and manufacturing to consumer use, disposal and any post-use loops of resell, refurbish and recycle. By highlighting the lifecycle inputs and outputs, the map draws attention to the fact that all products use raw materials derived from nature and that nature has to deal with any waste materials at the end of the life of a product. As such, the product lifecycle map shifts the focus from the end users or consumers to non-humans and non-users impacted as part of a product's lifecycle.

### *4.4. Behavioural impact canvas*

The behavioural impact canvas offers a structured approach to understanding, addressing, and potentially resolving complex environmental and social challenges by systematically analysing and transforming human behaviour (Clasen, 2023). Similar to other canvas design tools, such as the business model canvas (Osterwalder & Pigneur, 2010), it captures existing states and provides a structure for turning them into preferred future states. This is achieved through two steps: analysis and transformation. The analysis step involves dissecting the issue and linking it to harmful behaviour, motivations and external triggers and enablers. To that end, the analysis aims to uncover the problem's root causes, becoming the basis for the transformation step. The transformation step focuses on envisioning beneficial behaviours and finding solutions that support those behaviours. The tool was designed by one of the authors through her teaching and based on her experience as a design practitioner to provide valuable insights and practical solutions for understanding and improving human behaviour.

## 5. Findings

This section reports on our experience as first-person study participants regarding the implementation of the life-centred tools. For each tool, we report on how it helped us to better understand the subject (the Poäng chair) and what we learned about the tool. The observations draw on the reflections following the workshops. We use quotes from the workshop transcripts to illustrate specific findings.

### *5.1. Systems mapping*

The system mapping activity started with creating a mind map, which helped us see how the subject is part of a much bigger system (Figure 2). It was found to be a useful starting point for the team to get familiar with systems thinking, sharing knowledge and identifying gaps. However, the quality of the mind map greatly depends on existing knowledge within the team and previously collected data to feed into the activity. Specific to the chair, creating the mind map helped us think about the main materials that the chair was made of, where they came from and points of leverage for circular intervention. The mind map also allowed us to hypothesise about consumer behaviours, which was a useful foundation for subsequent tools. As one participant noted during the workshop: "Interestingly, regarding furniture, IKEA is currently working a lot on optimising for reassembly, because they realise that people move a lot, and if it's not possible to reassemble the furniture, then it doesn't get reused as much".

The systems map helped us to visualise how the chair is part of a much bigger system and highlighted the interconnectedness of components (Figure 3). The process became very involved very quickly, and without a clear boundary, it was hard to know where to stop. As one participant reflected, "It can get complicated when you look at every detail: from screws to fabric to wood". The concept of feedback loops was difficult to understand for those of us who had not used systems maps before. Attempting to identify feedback loops also made us realise that we were lacking domain expertise (e.g. from manufacturers and distributors). The systems map helped us to more systematically think through the chair's main materials as well as its lifecycle and impacts on other entities. This included revealing human and non-human actors and how their interconnectedness can be leveraged to consider life-centred interventions. For example, if the marketing department (human actor) advertises the chair as "highest selling", this increases sales, which in turn increase the extraction of raw materials (non-human actor).

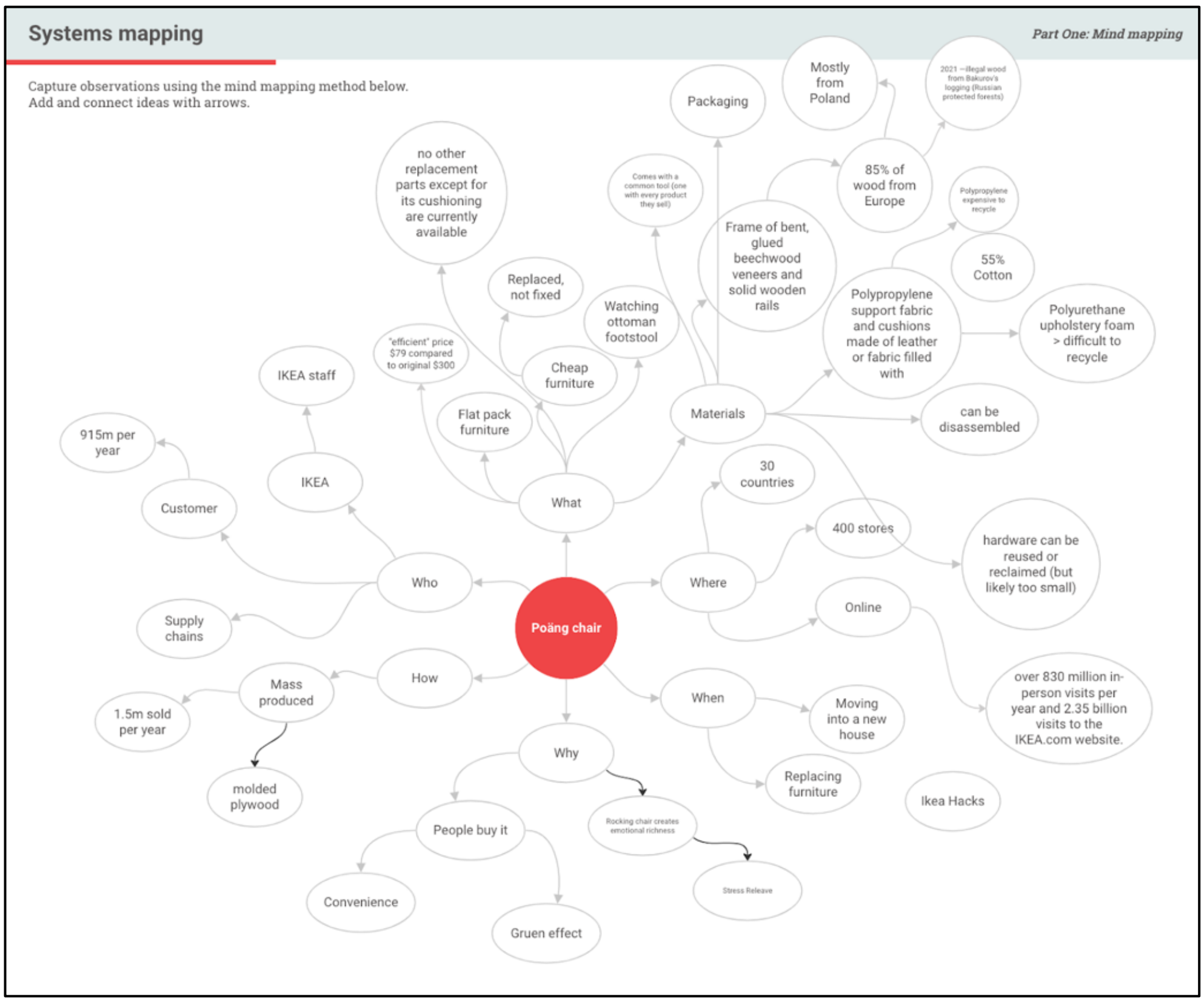

*Figure 2: The filled-in mind map for IKEA's Poäng chair as a preceding step for creating a systems map. The activity used a mind mapping variation that uses What, Why, How, Who, and When as prompts (Tomitsch, Borthwick, et al., 2021).*[1]

Creating mind maps and systems maps requires some knowledge of the subject, which in our case was based on secondary research but could be augmented through primary research or involving experts in the process as is the case for participatory systems mapping (Sedlacko et al., 2014). Both tools also require a clearly defined boundary to help participants understand when to stop and the level of detail that would be useful for the particular situation. Mind maps are an easy and accessible tool to get started, whereas systems maps require more explanation and practice to understand how to identify components and connections.

---

[1] A high resolution version of all the life-centred tools included in the paper can be accessed at: https://miro.com/app/board/uXjVNl6n4-o=/?share_link_id=735742248123

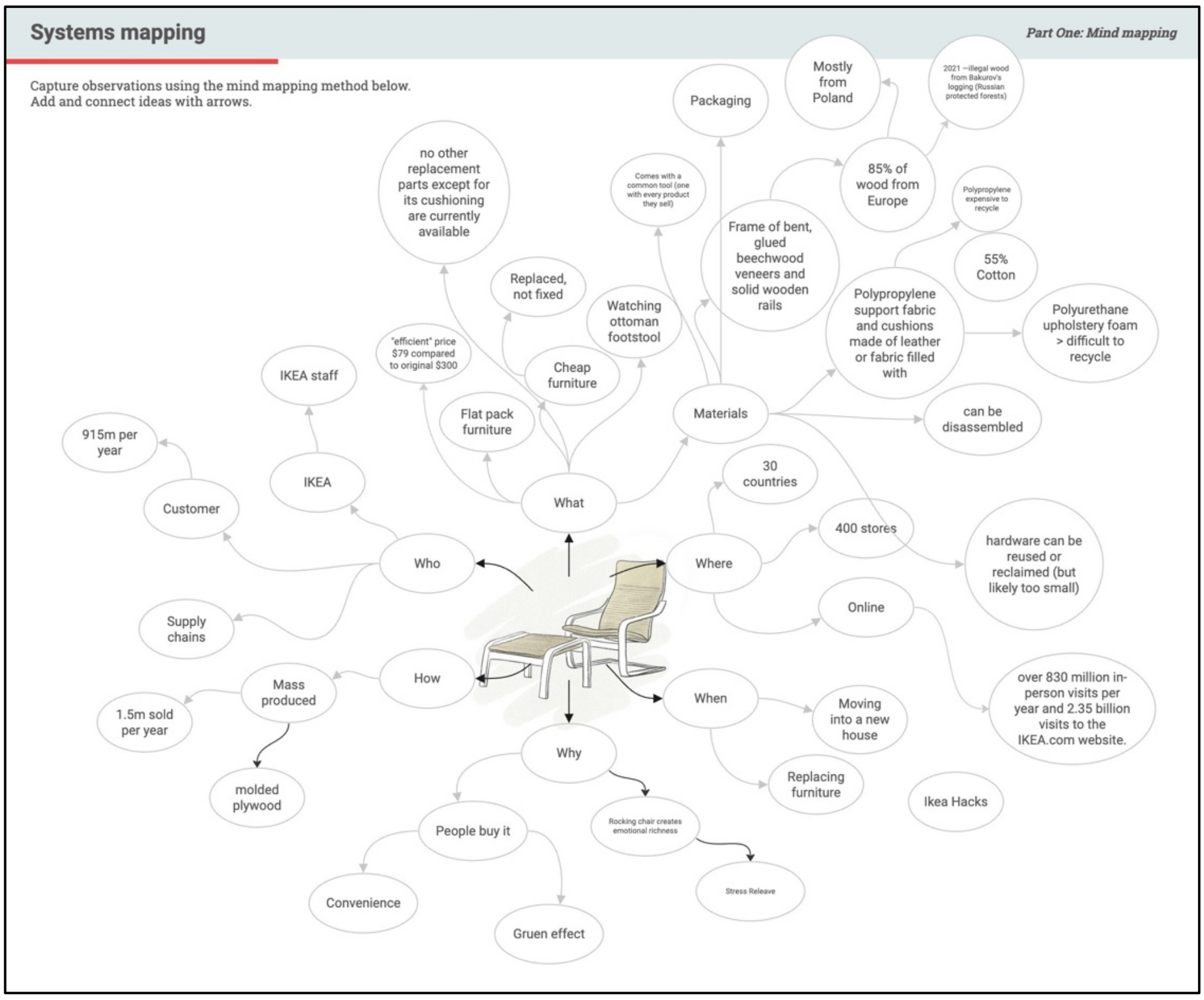

*Figure 3: The systems map (Tomitsch, Borthwick, et al., 2021) for the Poäng chair. The activity prompted participants to consider social, technological, economic, environmental and political components.*

## 5.2. Actant mapping

For creating the actant map, we drew on components from the systems map and categorised them into human and non-human actors (Figure 4). We found that we did not learn anything new about the chair itself through this tool, but it prompted us to identify other components that are part of the chair's systems. The tool helped us map out indirect, less obvious actors, such as the local council that has to collect and discard the chair at the end of its lifespan. Using different coloured sticky notes encouraged us to think about different categories of non-human actants, i.e. corporations and organisations; natural resources; policies, laws and regulations; and technologies. We found that filling in the map required a certain level of knowledge about the subject. As one participant pointed out, "In a real project, we wouldn't do this by ourselves, we wouldn't only do desk research but also invite others". However, going through the activity with our team provided a good starting point, helping us to find gaps and potential focus points. As another participant mentioned during the workshop, "We could try to talk to some IKEA business person or a logger but I don't know if that comes after, because how would you know that you need to talk to the logger if you don't do the mapping first?".

As a tool, the actant map was effective for identifying non-human actors and getting into systems thinking, which requires understanding the elements and interconnections that make up the content of systems (Arnold & Wade, 2015). As one of the participants reflected, the tool "gives an overview of the system". Compared to systems maps, we found actant maps to be less effective for determining points of intervention. However, they are a useful tool for identifying additional components that could subsequently be added to the systems map.

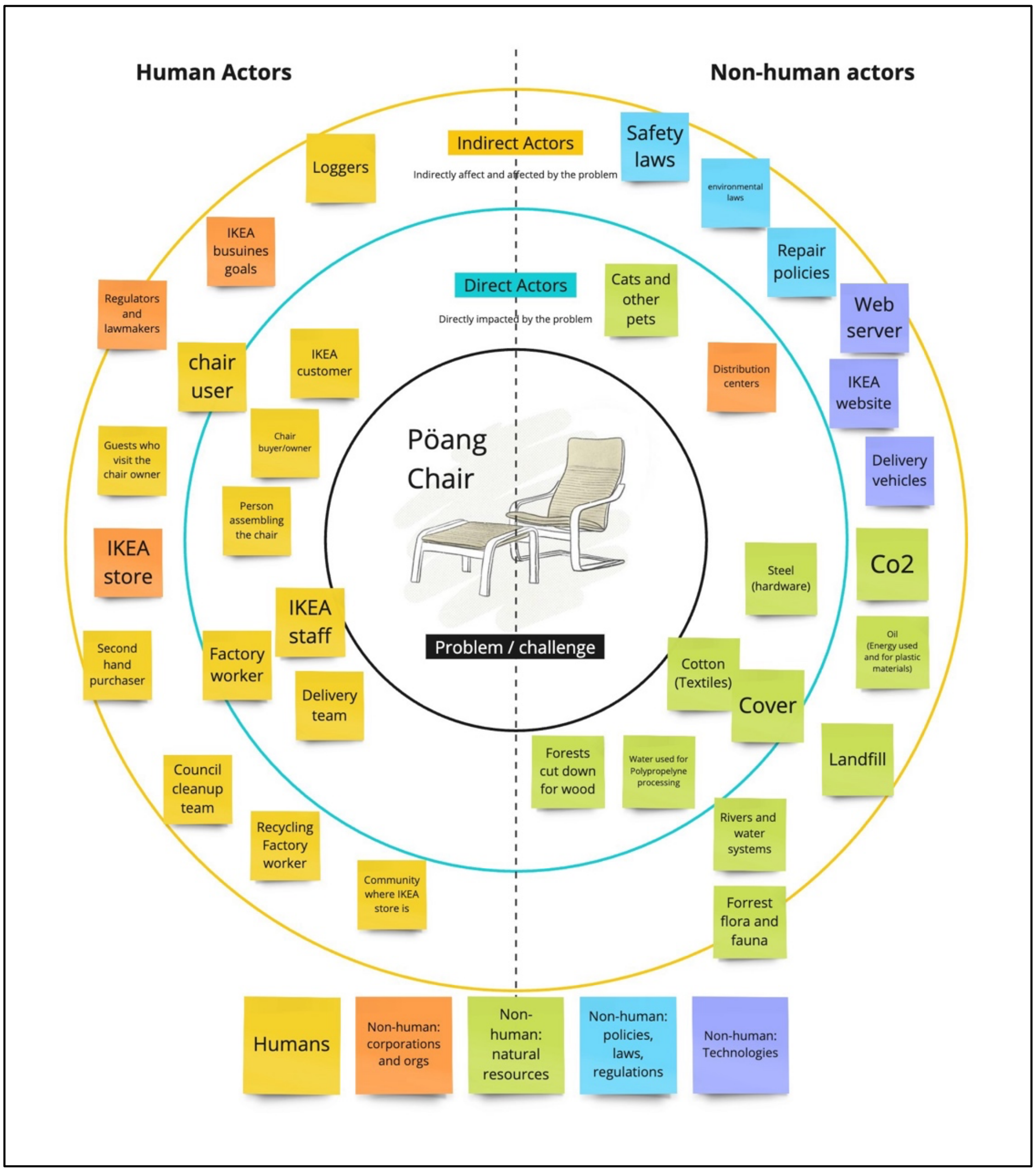

*Figure 4: The actant map (Sznel, 2020) for the Poäng chair. Concentric circles help visualise what/who is directly impacted by the problem or challenge and what/who is indirectly impacted or affecting the problem.*

## *5.3. Product lifecycle mapping*

To develop the product lifecycle map, we referenced the systems map and the mind map, looking for components that represented materials and natural resources (Figure 5). The tool enabled us to further build on the actant map and extend our understanding of the role of non-humans and non-users by linking them to the various phases of the product lifecycle. Specific to the Poäng chair, the tool revealed that no recycled materials were used in the chair's production (noting that we only had access to secondary research). The tool helped us identify where and how non-humans and non-users were impacted. Filling in the product lifecycle map required us to conduct additional secondary research, but in the process, we learned more about the manufacturing and material perspective. As one of our participants noted, "For me this what was an Aha! moment with this tool that it helps you with structuring, researching and like getting deeper into the nitty gritty of the thing".

The product lifecycle map provides a structured way to visualise the elements of a product lifecycle, from material extraction to end-of-use. Like the actant map, the product lifecycle map also contributes to developing a systems perspective by seeing the whole lifecycle and how it interacts with other elements. As a life-centred design tool, the map offers a way to identify intervention points, for example, considering how to reduce the reliance on natural resources and integrate recycled materials into the production process. To that end, the tool supports brainstorming opportunities for interventions; however, fully utilising its potential requires expert knowledge and time. As one participant reflected, "Trying to impact the whole life cycle is a big task". This observation further suggests that it might be helpful to follow up the product lifecycle mapping with another tool to scaffold the process of finding a focus for an intervention.

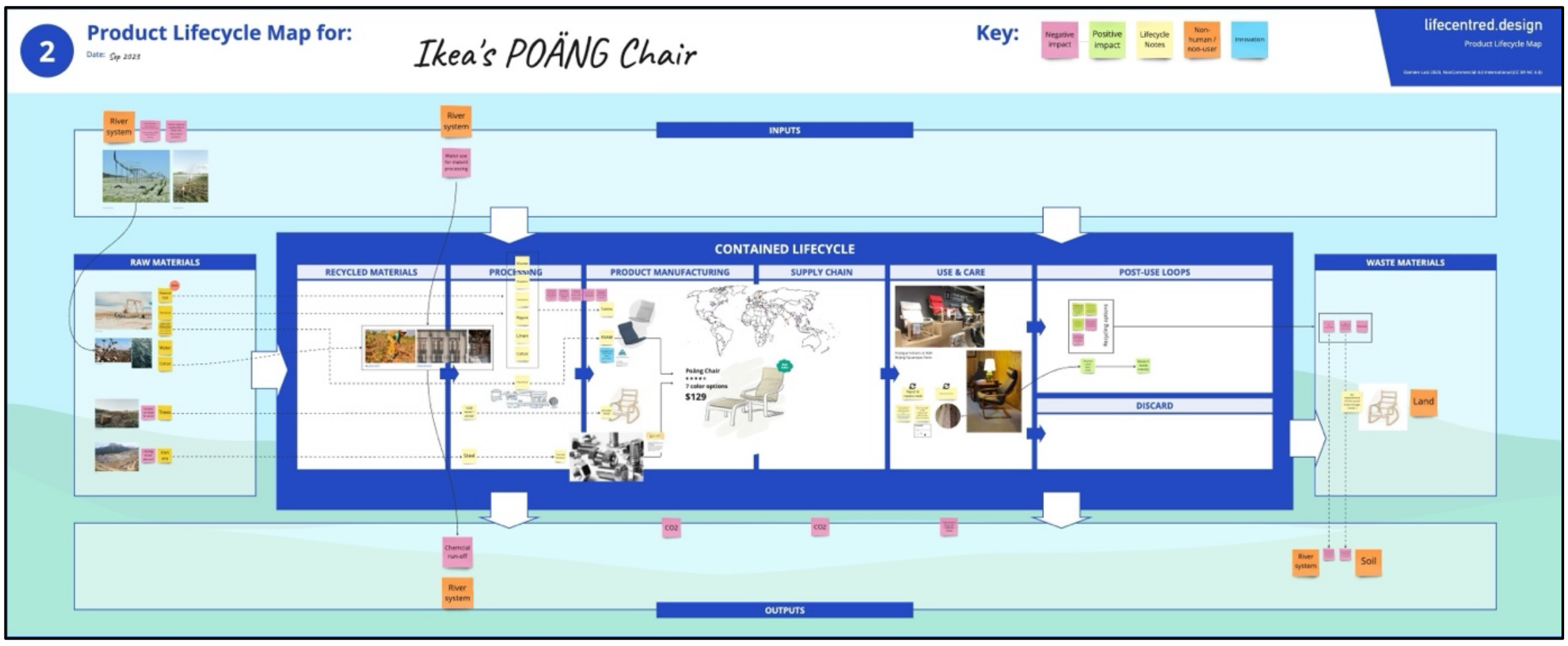

*Figure 5: Product lifecycle map (Lutz, 2023) for the Poäng chair. The map shows inputs, raw materials and outputs used for and created by each step in the lifecycle of a product, including waste.*

## *5.4. Behavioural impact canvas*

The behavioural impact canvas shifted the focus from the chair (as the subject) to human behaviour and how this behaviour is motivated by underlying models (such as the business model) and motivations (e.g., the need for affordable furniture that can be reassembled), demonstrated through an exchange between participants: one team member observed, "If we focus on buying the chair it could have a huge impact in the whole system because of the feedback loops"; another added "The company marketing the product is a behaviour as well I guess. Are we referring to the influence of marketing in consumer behaviour?"; and a third participant noted, "It's a human behaviour that people want affordable furniture that looks good."

Through its two-step approach, the canvas allowed us to systematically identify current behaviours (Figure 6) and subsequently consider how these behaviours could be transformed (Figure 7). As one participant reflected, "The tool was good to understand the relationship between motivation and impact". After each participant had filled in the canvas, we found ourselves building on each other's ideas (akin to other brainstorming processes (Kohn et al., 2011)), which triggered new ideas for potential interventions. This process led us to speculate on a model involving a free pickup service for furniture no longer needed, which is then refurbished and sold to another customer. As such, the tool has the potential to drive decisions towards life-centred design practices that consider the connection between aspects like the artefact and business models and how they impact the environment.

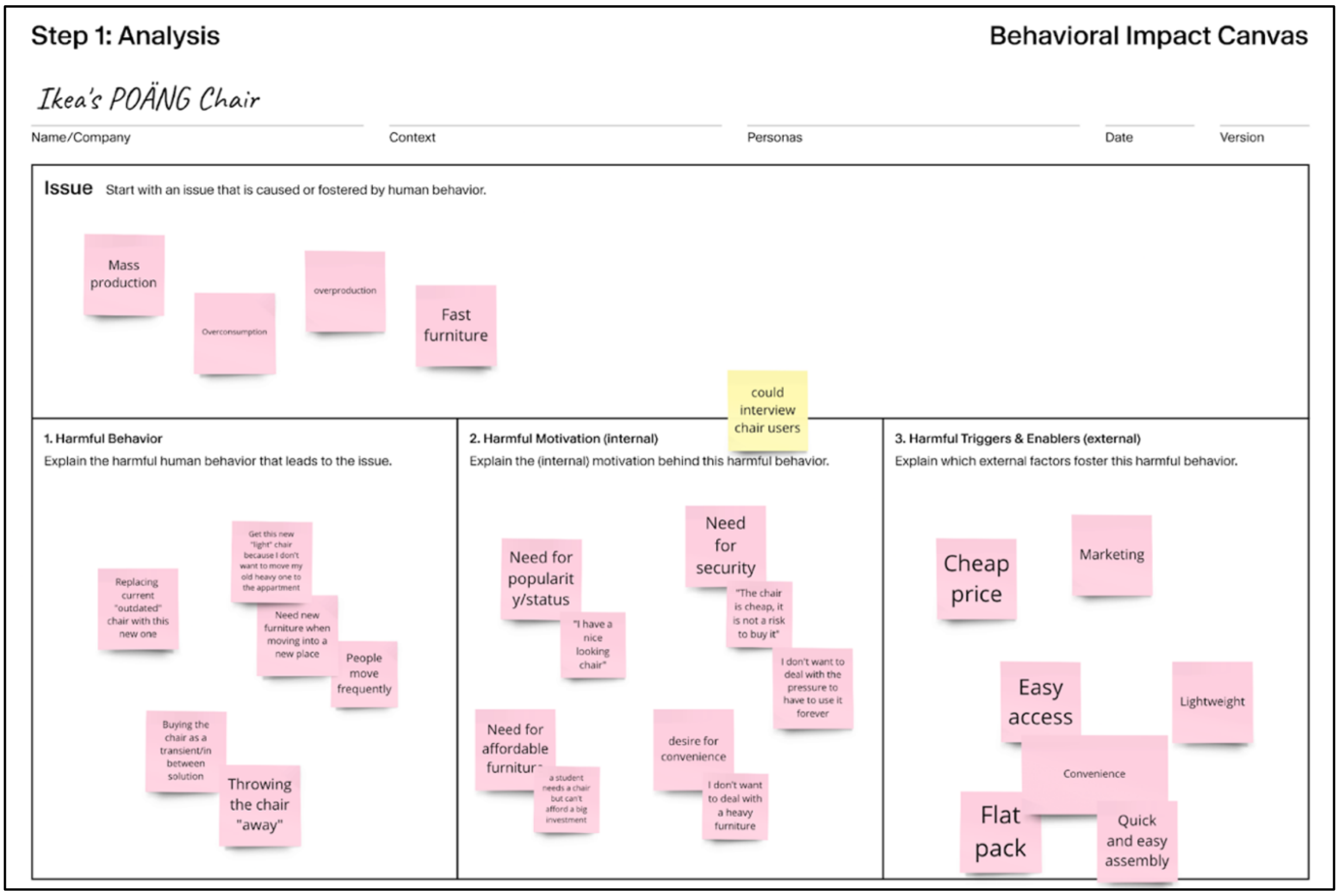

*Figure 6: The analysis step of the behavioural impact canvas (Clasen, 2023), capturing the human behaviours in relation to IKEA's Poäng chair, the motivations for these behaviours, and triggers and enablers.*

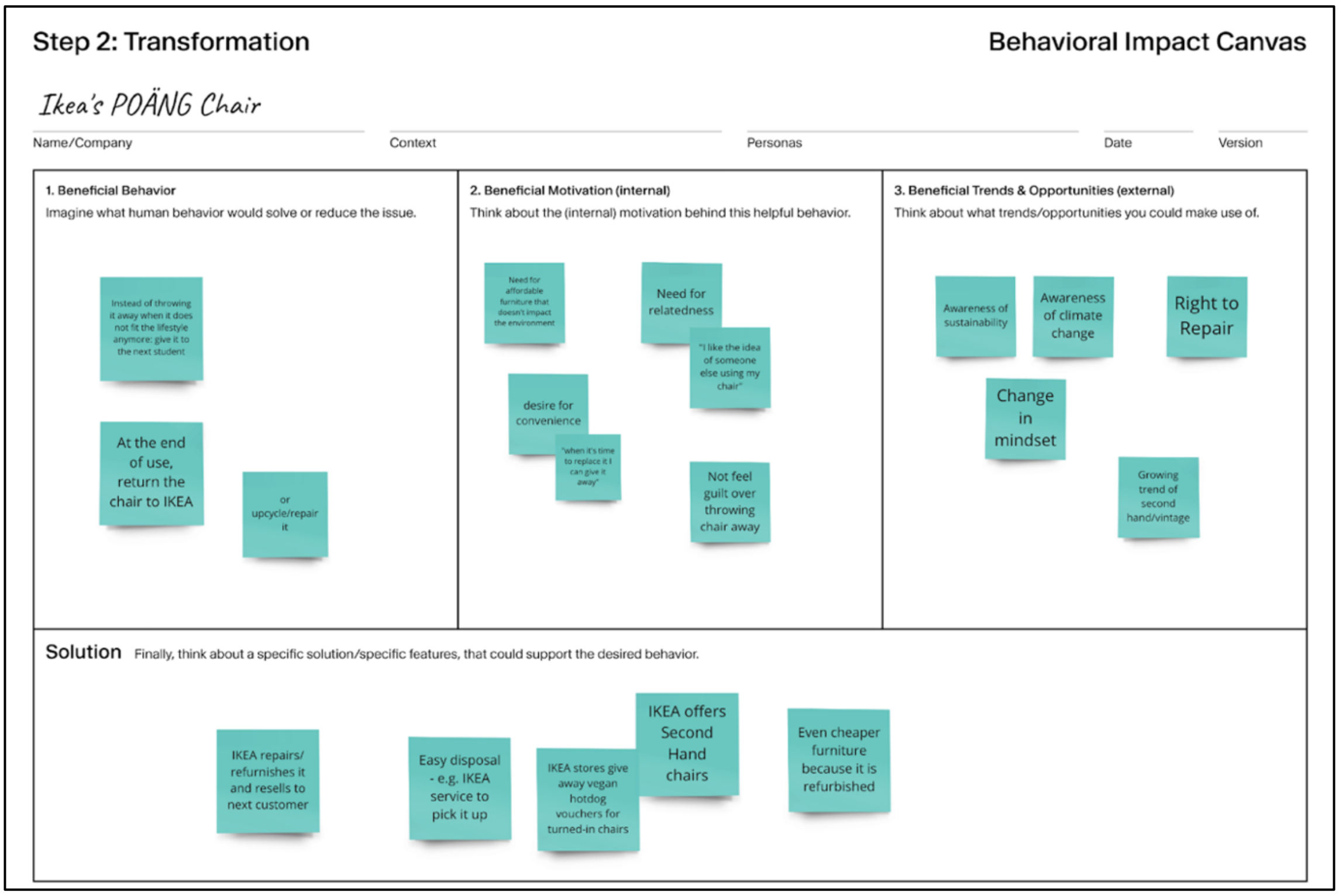

*Figure 7: The transformation step of the behavioural impact canvas (Clasen, 2023), turning the harmful behaviours into alternative positive behaviours that could drive sustainable solutions.*

While filling out the canvas, we noticed we were making assumptions about the behaviour, motivations, triggers and enablers that consumers currently exhibit. As a countermeasure, we went back to doing more secondary research, drawing on online articles that described the behavioural patterns of fast furniture consumers. However, the process would have benefitted from having primary research data available to draw on. On a philosophical level, the tool requires teams to consider what "beneficial" means – who benefits from the proposed intervention and in what ways? The answers to these questions influence the extent to which the tool can support designing for balance. As one participant reflected, while filling in the canvas, it's easy to "forget how the tool connects to non-humans, non-users and other life-centred perspectives". It is, therefore, important to set a foundation to guide the process – which in our case was achieved through preceding the behavioural impact canvas with other life-centred tools.

## 6. Discussion

Similar to human-centred design tools, the tools we used in our design study provided a structured way to work through a series of questions and goals. The difference lies in the objectives that shape how the tools are applied. The objective of human-centred design is to ensure design teams "satisfy the needs and wants of the user" when developing new products, services or systems (Maguire, 2001). Over the years, the remit of human-centred design has expanded to consider diverse perspectives (Krippendorff, 2004) and sustainability (DiSalvo et al., 2010), which is also expressed in the definition of human-centred design by the International Organization for Standardization (ISO, 2019). However, these aspects are often overlooked in practice and commonly lacking in human-centred tools. One of the applications of life-centred design is to redesign human-centred artefacts. This motivated our choice of the Poäng chair, although it also limited the potential to advocate for a life-centred approach. At the same time, as Dilnot put it, "the only way out of the anthropocentric (out of crisis) is through the anthropocentric" (Dilnot, 2021, p.55).

## *6.1 Usefulness and limitations of the selected tools*

Reflecting on the usefulness and limitations of the selected tools, we make a number of observations.

*Focusing on systems and non-humans/non-users.* The objectives of life-centred design tools are to design with all life's needs in mind, for all human pluriversal diversity, and for a symbiotic relationship with the impacted natural and human sources (Lutz, 2022). The tools we used in this study – systems mapping, actant mapping, product lifecycle mapping and behavioural impact canvas – support achieving these objectives through a focus on systems thinking and prompting us to consider non-humans and non-users. Would our initial focus have been on systems thinking instead of taking a life-centred approach, we would have likely ended up with a different set of tools and a different outcome. Life-centred tools have the ability to translate principles of Actor-Network Theory into design practice, implementing Latour's call to reject the distinction between nature and culture and recognise society as a complex web of interconnected actors (Latour, 2005).

*Involving domain experts and decision-makers.* Human-centred tools have been criticised for being prone to bias as designers choose to represent some data points over others, for example, when creating persona representations (Marsden & Haag, 2016). Life-centred tools are not immune to this risk. Indeed, at several stages during our study, we found ourselves making assumptions when implementing the tools. This highlights that being aware of potential biases and assumptions is just as important as fluency in using the tools. To reduce bias and increase the validity of data represented in the tools, designers can involve domain experts and people with lived experience. In the case of non-humans, their voice can be brought into the process through creating a coalition (Tomitsch, Fredericks, et al., 2021). Our team's practice-based members also highlighted the importance of getting buy-in. As one participant noted during the second product lifecycle mapping session, "I feel like you need to bring people along from the beginning in order to like sell anything to anybody in the company". This highlights the strategic value of involving key stakeholders in life-centred tools to drive positive change within an organisation.

*Identifying opportunities as well as gaps.* We found that even implementing each of the tools without experts and other participants in the first instance was effective in expanding our understanding of the subject and potential opportunities for interventions. The process also helped us to identify gaps in our knowledge, highlighting areas for further investigation.

*Selecting and sequencing the tools.* The order in which we used the tools enabled us to move from general, broad concerns to more specific, focused investigations. Based on our findings, we suggest a slightly adjusted order and inserting a step for deciding on a focus (Figure 8). The behavioural impact canvas could be followed by creating an impact ripple canvas (Tomitsch, Borthwick, et al., 2021) (to assess potential interventions for unintended consequences) and a triple-layered business model canvas (Joyce & Paquin, 2016) (to brainstorm alternative more balanced business models). Critically, part of the process should be for designers to understand why they are using the tools, for what goals and how the tools are used to achieve those goals.

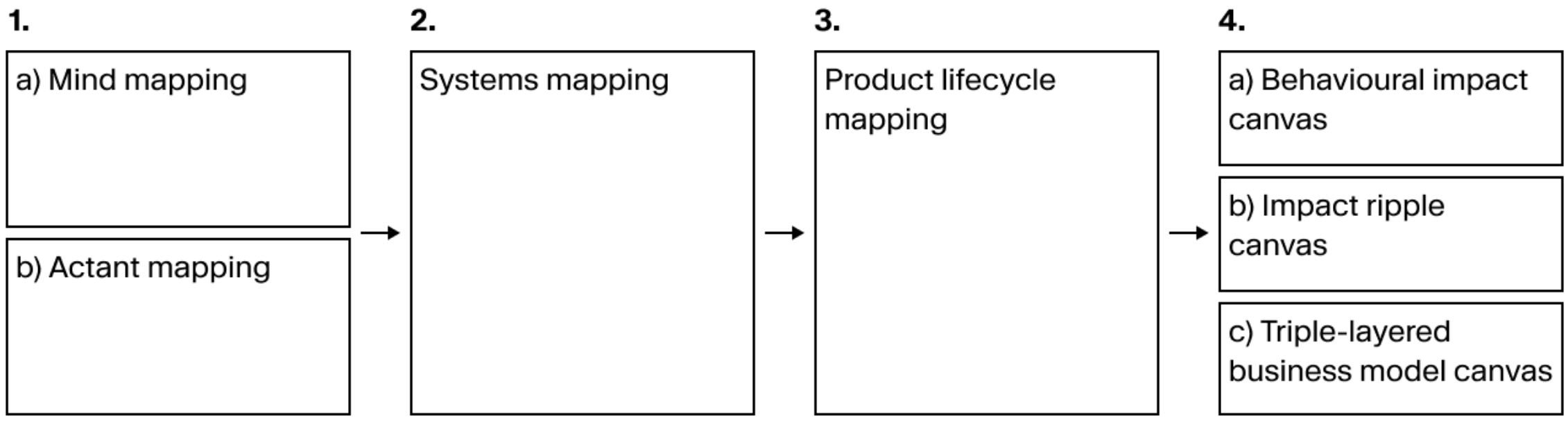

*Figure 8: Potential order for implementing the tested life-centred design tools (1-4a) in design projects and suggestions for what tools to use in a subsequent step (4b-c).*

### *6.2 Study limitations*

The validity of our findings is constrained by the limits of our perspectives as authors and participants. Although we brought together the cultural perspectives from three continents and across academia and practice, our positionality still drove the selection of the tools and what data we chose to include or not include. As this used a hypothetical brief, we did not involve experts from other domains, such as material science or supply chain management. Two of the paper authors were also the authors of the books that served as a basis for choosing some of the methods. This added an expert lens but also potential biases as to the selection of tools. By selecting tools, we implicitly disregarded others that may provide value when designing for balance. As such, the account of the life-centred tools, order of use and which ones would be needed in a process is demonstrative and intended to serve as a starting point. Future studies could conduct a more systematic review of a wider range of tools and their contribution to developing life-centred design mindsets.

## 7. Conclusion

Through reflecting on the findings from a first-person design study, the paper investigated life-centred design as a process and a set of tools to support the shift towards designing for systemic balance. Responding to the DRS theme track “Design for Balance: Reimagining Pro-cesses and Competences for Sustainable Futures", the paper specifically interrogated challenges associated with a human-centricity as it is commonly applied in practice. The paper highlights the role of life-centred design as a new competency for sustainable futures as an extension of human-centred design practice. The tools and process for how to implement the tools offer a starting point for adopting this approach in practice and supporting the development of this competency in design education.

## About the Authors:

**Martin Tomitsch** leads the Transdisciplinary School at the University of Technology Sydney. Through his work as design academic and educator, he advocates for the transformative role of design as a framework to imagine speculative futures and drive positive change.

**Katharina Clasen** is a Senior UX Designer and lecturer from Germany. With the goal to bring life-centricity to the design practice, she founded LifeCenteredDesign.Net, co-founded the German UPA working group *Design for Sustainability* and developed the Behavioral Impact Canvas.

**Estela Duhart** is a Mexican Service Designer and Researcher based in Chicago, working towards a more equitable, just, and life-centered design practice using frameworks like Biomimicry and Design Justice Principles.

**Damien Lutz** is a Sydney-based UX Designer and life-centred design consultant. With a personal mission to evolve design to be more kind and regenerative, Damien developed the first life-centred design framework and toolkit and founded the lifecentred.design online hub.